\documentclass[superscriptaddress,twocolumn,aps,pra,showpacs,
fixfloats]{revtex4-1}

\usepackage{bm}
\usepackage{dcolumn}
\usepackage{graphicx}
\begin{document}

\title{Electron elastic scattering off $A$@C$_{60}$: The role of atomic polarization under confinement}

\author{V. K. Dolmatov}
\affiliation{University of North Alabama,
Florence, Alabama 35632, USA}
\author{ M. Ya. Amusia}
\affiliation{Racah Institute of Physics, Hebrew University, 91904 Jerusalem, Israel}
\affiliation{A. F. Ioffe Physical-Technical Institute, 194021 St. Petersburg, Russia }
\author{L. V. Chernysheva$^{3}$}

 \date{\today}

\begin{abstract}
The present paper explores possible features of electron elastic scattering off endohedral fullerenes $A$@C$_{60}$. It focuses on how dynamical polarization of the encapsulated atom $A$ by an incident electron might
alter scattering off $A$@C$_{60}$ compared
to the static-atom-$A$ case, as well as how the C$_{60}$ confinement modifies the impact of atomic polarization on electron scattering compared to the free-atom case. The aim is to provide researchers with a ``relative frame of reference'' for understanding which part of the scattering processes could be due to electron scattering off the encapsulated atom and which due to scattering off the C$_{60}$ cage.
To meet the goal, the C$_{60}$ cage is modeled by an attractive spherical potential of a certain inner radius, thickness, and depth which is a model used frequently in a great variety of fullerene studies to date. Then, the Dyson equation for the self-energy part of the Green's function
of an incident electron moving in the combined field of an encapsulated atom $A$ and C$_{60}$ is solved in order to account for the impact of dynamical polarization of the encaged atom upon $e + A@{\rm C_{60}}$ scattering. The Ba@C$_{60}$ endohedral is chosen as the case study.
The impact is found to be significant, and its utterly different role compared to that in $e + Ba$ scattering is unraveled.
\end{abstract}


\maketitle

\section{Introduction}
Electron elastic scattering off quantum targets is an important fundamental phenomenon of
nature. It has significance to both the basic and applied sciences
and technologies. Yet, to date, the knowledge on the process of
electron collision with such important quantum
targets as endohedral fullerenes $A$@C$_{60}$ is far from complete.
Endohedral fullerenes are
nanostructure formations where an atom $A$ is encapsulated
inside the hollow interior of a C$_{60}$ molecule. The authors are aware of only a handful of works on this
subject. These are the theoretical studies of fast charged-particle ionization of $A$@C$_{60}$ \cite{Balt09,Amusia11,Cruz13} and
low-energy electron scattering  off $A$@C$_{60}$ \cite{DolmJPB,DolmPRA,AmCherJETPlett}
calculated in the framework of two different model approximations. Namely, in Refs.~\cite{DolmJPB,DolmPRA} a static Hartree-Fock (HF) approximation was employed. There, both the atom $A$ and C$_{60}$ were considered as
 nonpolarizable targets and  the C$_{60}$ cage was modeled by an attractive spherical potential of a certain inner radius, thickness, and depth. In Ref.~\cite{AmCherJETPlett},
 the authors kept the encaged atom ``frozen'', modeled the C$_{60}$ cage by the potential similar to that used in Ref.~\cite{DolmJPB,DolmPRA}, but accounted for polarization of the C$_{60}$ cage by incident electrons. The latter was evaluated
 in a simplified manner by adding a static polarization potential $-\alpha/r^{4}$ ($\alpha$ being the static polarizability of C$_{60}$) to the model C$_{60}$-potential. Note, a meager amount of research on
  $e +A@{\rm C_{60}}$ collision is in contrast to the study of photon-$A$@C$_{60}$ collision, different aspect of which has been intensely
scrutinized in a great variety of theoretical works to date (see, e.g., Refs.~\cite{Baltenkov99,JPCVKDSTM99,Gorczyca12,Comment,Amusia14,O'Sullivan13,@TD,Himadri14} and references therein),
 including experimental studies \cite{Phaneuf10, Phaneuf13} (and references therein). Such disbalance in favor of the number and quality of
studies of $A$@C$_{60}$ photoionization versus research on electron-$A$@C$_{60}$ collision is not accidental. Electronic collision  with
a multielectron target is a more complicated multifaceted process compared to photonic collision with the same target.
Therefore, the comprehensive description
of electron scattering by a multielectron target is too challenging
for theorists even with regard to a free atom, not to mention $A$@C$_{60}$ targets.

The present study does not aim at solving the difficult problem of electron-$A$@C$_{60}$ scattering in its entirety. Instead, it focuses on the contribution of electron scattering only off the encapsulated atom $A$ to the entire collision process. The significance of the present study is that it provides an important frame of reference for (future) understanding of which part of electron-$A$@C$_{60}$ scattering could be due to scattering off the encapsulated atom $A$ (unraveled in the present work) and which is due to other ``facets'' of the entire $A$@C$_{60}$ system. Research results, thus, provide a relative rather than absolute knowledge. To meet the goal, the C$_{60}$ cage is modeled, as in Refs.~\cite{DolmJPB,DolmPRA,AmCherJETPlett,Winstead}, by a spherical potential of a certain inner radius, thickness, and depth. Polarization of the C$_{60}$ cage by incident electrons will, thus, be ignored (being not the subject of the focused study). This is in contrast to accounting for polarization of the encapsulated atom in the present study. The neglect by polarizability of C$_{60}$ by incident electrons should not be over-dramatized. The effect of polarizability is electron-energy-dependent and may either enhance or decrease scattering cross section, at certain electron energies. Therefore, when scattering off C$_{60}$ is dramatically decreased, or where scattering of the encapsulated atom $A$ is dramatically increased, a relative role of scattering off the atom $A$ will (might) be significant. Furthermore, $A$@C$_{60}$ has the hollow interior which is not totally occupied by the atom (i.e., not totally filled in with charge density). As such, it acts as a resonator relative to incident electronic waves. Therefore, at wave frequencies, matching resonance frequencies of the $A$@C$_{60}$-resonator, there will be a significant incident-electron-density build-up in the hollow interior of $A$@C$_{60}$. Obviously, this build-up of electron density will be positioned near the encapsulated atom $A$. Therefore, the effect of atomic polarization on electron scattering might become comparable or even more important than the C$_{60}$ polarization effect, at resonance frequencies.  As such, the impact of atomic polarizability on $e +A@{\rm C_{60}}$  scattering it cannot be dropped out of the consideration at all. It is, therefore, indisputably needed (and interesting, and does make sense) to study how polarization of the encapsulated atom can affect the scattering process even in the neglect by polarization of the fullerene cage by incident electrons. To account for atomic polarization under confinement, the authors employ the Dyson formalism
for the self-energy part of the Green's function of a scattered electron \cite{Abrikosov,ATOM}, adapt it to the case of the electron motion in a combined field of the encapsulated multielectron atom $A$ and the model static C$_{60}$ cage, solve
the generalized Dyson equation, and, thus, calculate the electron elastic-scattering phase shifts and corresponding cross sections for the $e + A@{\rm C_{60}}$ scattering reaction. The study is restricted to electron elastic scattering at low electron energies $\epsilon \alt 3$ eV where the most interesting effects occur.

Finally, the present study also has a significance which is independent of its direct applicability to electron-$A$@C$_{60}$ scattering. This is because it falls into a mainstream of intense modern studies where numerous aspects of the structure and spectra of atoms under various kinds of confinements (impenetrable spherical, spheroidal, dihedral, Debye-like potentials, etc.)  are being attacked from many different angles by
research teams world-wide (see, e.g., numerous review articles in Refs.~\cite{AQC57,AQC58,Kalidas14}). Such studies are interesting from the view point of basic science.
Results of the present study add new knowledge to the collection of atomic properties under confinement as well, particularly revealing the impact of atomic polarization under confinement on electron-atom scattering.

Atomic units are used throughout the paper unless specified otherwise.

\section{Theory}
\subsection{$e + A@{\rm C}_{60}$ scattering in the framework of static C$_{60}$}
\subsubsection{Model static HF approximation}

In the present work, the C$_{60}$ cage is modeled by a spherical potential $U_{\rm c}(r)$ defined as follows:
\begin{eqnarray}
U_{\rm c}(r)=\left\{\matrix {
-U_{0}, & \mbox{if $r_{0} \le r \le r_{0}+\Delta$} \nonumber \\
0 & \mbox{otherwise.} } \right.
\label{SWP}
\end{eqnarray}
Here, $r_{0}$, $\Delta$, and $U_{0}$ are the inner radius, thickness, and depth of the potential well, respectively.

Next, the wavefunctions $\psi_{n \ell m_{\ell} m_{s}}({\bm r}, \sigma)=r^{-1}P_{nl}(r)Y_{l m_{\ell}}(\theta, \phi) \chi_{m_{s}}(\sigma)$
and binding energies $\epsilon_{n l}$ of atomic electrons
($n$, $\ell$,  $m_{\ell}$ and $m_{s}$ is the standard set of quantum numbers of an electron in a central field, $\sigma$ is the electron spin variable) are the solutions of a system of the ``endohedral''
HF equations:
\begin{eqnarray}
&&\left[ -\frac{\Delta}{2} - \frac{Z}{r} +U_{\rm c}(r) \right]\psi_{i}
({\bm x}) + \sum_{j=1}^{Z} \int{\frac{\psi^{*}_{j}({\bm x'})}{|{\bm
x}-{\bm x'}|}} \nonumber \\
 && \times[\psi_{j}({\bm x'})\psi_{i}({\bm x})
- \psi_{i}({\bm x'})\psi_{j}({\bm x})]d {\bm x'} =
\epsilon_{i}\psi_{i}({\bm x}).
\label{eqHF}
\end{eqnarray}
Here, $Z$ is the nuclear charge of the atom, ${\bm x} \equiv ({\bm r}, \sigma)$, and the integration over ${\bm x}$ implies both the integration over ${\bm r}$ and summation over
$\sigma$. Eq.~(\ref{eqHF}) differs from the ordinary HF equation for a free atom by the presence of the $U_{\rm c}(r)$ potential in there. This equation is first solved in order to calculate the electronic ground-state wavefunctions of the encapsulated atom. Once the electronic ground-state wavefunctions are determined, they are plugged back into
 Eq.~(\ref{eqHF}) in place of $\psi_{j}({\bm x'})$ and $\psi_{j}({\bm x})$ in order to calculate the electronic wavefunctions of scattering-states $\psi_{i}({\bm x})$ and their radial parts $P_{\epsilon_{i}\ell_{i}}(r)$.

 Corresponding electron elastic-scattering phase shifts $\delta_{\ell}(k)$
  are then determined by
 referring to $P_{k\ell}(r)$ at large $r$:
\begin{eqnarray}
P_{k\ell}(r) \rightarrow \sqrt{\frac{2}{\pi}}\sin\left(k r -\frac{\pi\ell}{2}+\delta_{\ell}(k)\right).
\label{P(r)}
\end{eqnarray}
Here, $k$ and $k'$ are the wavenumbers of the incident and scattered electrons, respectively, and
$P_{k\ell}(r)$ is normalized to $\delta(k-k')$.
The total electron elastic-scattering cross section $\sigma_{\rm el}(\epsilon)$ is then found in accordance with the standard
formula for electron scattering by a central-potential field:
 \begin{eqnarray}
 \sigma_{\rm el}(k)= \frac{4\pi}{k^2}\sum^{\infty}_{\ell=0}(2\ell+1)\sin^{2}\delta_{\ell}(k).
 \label{sigma}
 \end{eqnarray}

This approach solves the problem of $e + $A$@{\rm C_{60}}$ in a static approximation, i.e.,
without accounting for polarization of the $A$@C$_{60}$ system by incident electrons.

\begin{figure}[tbp]
\includegraphics[width=\columnwidth]{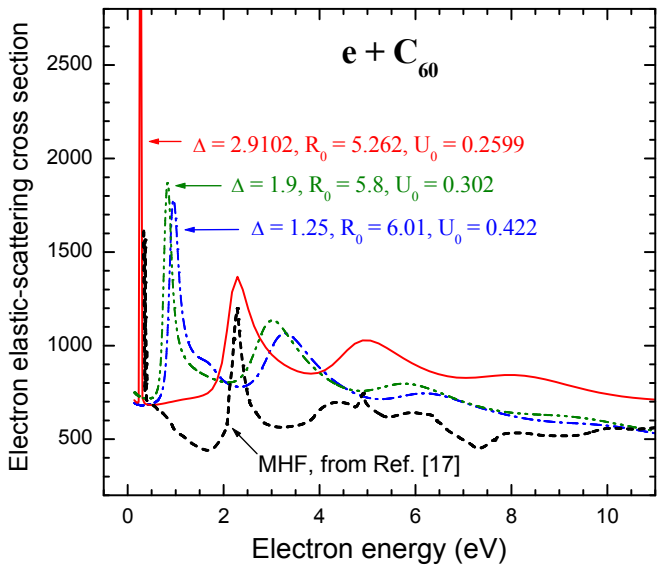}
\caption{(Color online) $e + {\rm C_{60}}$ elastic-scattering cross section (in units of $a_{0}^{2}$, $a_{0}$ being the first Bohr radius of the hydrogen atom)
calculated both with the use of different values of the parameters
$r_{0}$, $\Delta$, and $U_{0}$ of the
spherical potential $U_{\rm c}(r)$ (present work) and in the framework of {\em ab initio} MHF \cite{Winstead}, as marked.}
\label{C60}
\end{figure}

In the literature, some inconsistency is present in
choosing the magnitudes of $\Delta$, $U_{0}$ and $r_{0}$ of the model potential $U_{\rm c}(r)$ for C$_{60}$:
$r_{0}=5.8$, $\Delta=1.9$, $U_{0}=0.302$ \cite{JPCVKDSTM99,O'Sullivan13} (and references therein),
or $r_{0}= 6.01$, $\Delta=1.25$ and $U_{0}=0.422$ \cite{Amusia11,Gorczyca12}, or  $\Delta = 2.9102$, $r_{0} = 5.262$, $U_{0} = 0.2599$
\cite{Winstead}. In order to evaluate which of these sets of parameters is best suited for studying $e + A@{\rm C_{60}}$ scattering,
we performed the corresponding calculations of $e + {\rm C_{60}}$ scattering. Calculated results are plotted in Fig.~\ref{C60} against
calculated data obtained in the framework of the sophisticated \textit{ab initio} static-exchange molecular-Hartree-Fock (MHF) approximation \cite{Winstead}.

One can see that it is the set of parameters proposed in Ref.~\cite{Winstead} which leads \cite{Winstead} to the overall qualitative and
semi-quantitative agreement between some of the most prominent features of
$e + {\rm C_{60}}$ elastic scattering predicted by the model spherical-potential approximation and \textit{ ab initio} MHF.
Correspondingly, in the present work, as in Ref.~\cite{Winstead}, the $U_{\rm c}(r)$ potential is defined by
$\Delta = 2.9102$, $r_{0} = 5.262$, and $U_{0} = 0.2599$.

\subsubsection{Multielectron approximation: a polarizable atom $A$}

In order to account for the impact of polarization
of an encapsulated atom $A$ by incident electrons on $e + A@{\rm C}_{60}$
elastic scattering, let us utilize the concept of the self-energy part of the
 Green's function of an incident electron \cite{Abrikosov,ATOM}.

\begin{figure}[tbp]
\includegraphics[width=\columnwidth]{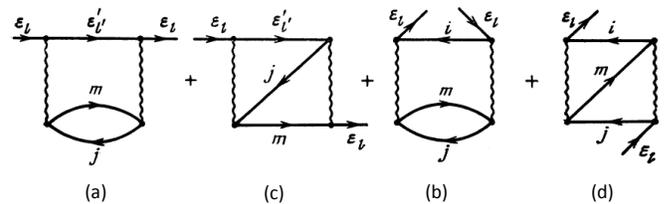}
\caption{The irreducible self-energy part $\Sigma(\epsilon)$
of the Green function of a
scattering electron in the second-order perturbation theory in the Coulomb interaction, referred to as the SHIFT approximation (see text). Here,
a line with a right arrow denotes an electron, whether a scattered electron
(states $|\epsilon_{\ell}\rangle$ and $|\epsilon_{\ell^{\prime}}^{\prime}\rangle$) or an atomic excited electron (a state $|m \rangle$), a line with a
left arrow denotes a vacancy (hole) in the atom (states $\langle j|$ and $\langle i|$), a
wavy line denotes the Coulomb interelectron interaction $V$.}
\label{SHIFT}
\end{figure}

In the simplest second-order perturbation theory in the Coulomb
interelectron interaction $V$ between the incident and atomic electrons,
the  \textit{irreducible} self-energy part of the
Green's function $\Sigma(\epsilon)$ of the  incident  electron
 is depicted with the help of Feynman diagrams in Fig.~\ref{SHIFT}.

The diagrams of Fig.~\ref{SHIFT} illustrate how a scattered electron ``$\epsilon _{\ell }$'' perturbs (read: polarizes) a $j$-subshell
of the atom by causing $j$ $\rightarrow $ $m$ excitations from the subshell and then gets coupled with
these excited states itself via both the Coulomb direct [diagrams (a) and (b)] and exchange [diagrams (c) and (d)] interactions.
Numerical calculation of electron elastic-scattering phase shifts in the framework of this approximation is addressed by the computer code from Ref.~\cite{ATOM} labeled as the ``SHIFT'' code.
Correspondingly, the authors refer to this approximation as the ``SHIFT'' approximation everywhere in the present paper.

\begin{figure}[tbp]
\includegraphics[width=\columnwidth]{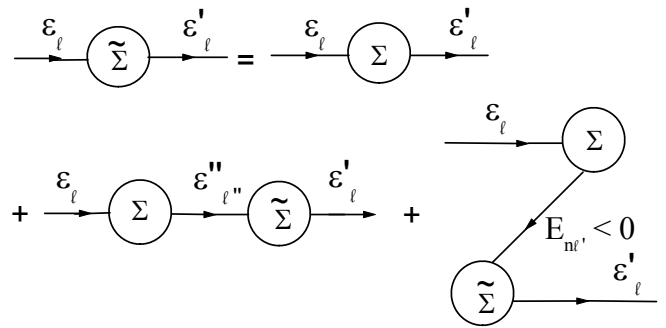}
\caption{The matrix element of the reducible self-energy part $\tilde{\Sigma}(\epsilon)$
of the Green's function of a scattering electron, where $\Sigma$ is the irreducible self-energy part of the Green's function depicted in Fig.~\ref{SHIFT}. This approximation is
referred to as the SCAT approximation (see text). Note, when calculating
$\langle \epsilon_{\ell}|\tilde{\Sigma}|\epsilon_{\ell}\rangle$ analytically, the summation over unoccupied discreet and integration over continuum excited states (marked as $\epsilon_{\ell ''}''$)
along with the summation over the occupied states in the atom marked as $E_{n\ell '}$ must be performed.}
\label{SCAT}
\end{figure}

A fuller account of electron correlation (read: polarization) in $e + A@{\rm C}_{60}$ elastic scattering is determined by the \textit{reducible}
$\tilde{\Sigma}(\epsilon)$ part of the self-energy part of
the electron's Green function \cite{ATOM}. The matrix element of the latter are represented diagrammatically in Fig.~\ref{SCAT}.

The above diagrammatic equation for $\tilde{\Sigma}$ can be written in an operator form as follows:
\begin{eqnarray}
\hat{\tilde{\Sigma}}=\hat{\Sigma}-\hat{\Sigma}\hat{G}^{(0)}\hat{\tilde{\Sigma}}
\label{EqGreen}.
\end{eqnarray}
Here, $\hat{\Sigma}$
is the operator of the  irreducible
self-energy part of the Green's function calculated in the
framework of SHIFT (Fig.~\ref{SHIFT}), $\hat{G}^{(0)}=(\hat{H}^{(0)}-\epsilon)^{-1}$ is the HF operator of the
electron Green's function, and $\hat{H}^{(0)}$ is the HF
Hamiltonian operator of the $electron + A@C_{60}$ system. Clearly, the equation for the matrix elements of $\tilde{\Sigma}$
accounts for an infinite series of diagrams by coupling the diagrams of Fig.~\ref{SHIFT} in various combinations.
Numerical calculation of electron elastic-scattering phase shifts in the framework of this approximation is addressed by the computer code from Ref.~\cite{ATOM} labeled as the ``SCAT'' code.
Correspondingly, the authors refer to this approximation as the ``SCAT'' approximation everywhere in the present paper.
SCAT works well for the case of electron scattering off free atoms \cite{ATOM}.
This gives us confidence in that SCAT is a sufficient approximation for pinpointing the impact of correlation/polarization
on $e + A@{\rm C_{60}}$ scattering as well.

In the framework of SHIFT or SCAT, the electron elastic-scattering
phase shifts $\zeta_{\ell}$  are determined as follows \cite{ATOM}:
\begin{eqnarray}
\zeta_{\ell }=\delta_{\ell }^{\rm HF}+\Delta\delta_{\ell }.
\end{eqnarray}
Here, $\Delta\delta_{\ell}$ is the correlation/polarization
correction term to the calculated HF phase shift $\delta_{\ell}^{\rm HF}$:
\begin{eqnarray}
\Delta\delta_{\ell}=\tan^{-1}\left(-\pi
\left\langle\epsilon\ell|\tilde{\Sigma}|\epsilon\ell\right\rangle \right).
\end{eqnarray}
The mathematical expression for
$\left\langle\epsilon\ell|\tilde{\Sigma}|\epsilon\ell\right\rangle$
is cumbersome. The
interested reader is referred to \cite{ATOM} for details. The matrix element
$\left\langle\epsilon\ell|\tilde{\Sigma}|\epsilon\ell\right\rangle$ becomes
complex for electron energies exceeding the
ionization potential of the atom-target, and so does the correlation term
$\Delta\delta_{\ell}$ and, thus, the phase shift
 $\zeta_{\ell}$ as well. Correspondingly,
\begin{eqnarray}
\zeta_{\ell }=\delta_{\ell}+i\mu_{\ell},
\end{eqnarray}
where
\begin{eqnarray}
\delta_{\ell}=\delta_{\ell }^{\rm HF}+
\rm Re\Delta\delta_{\ell},\quad \mu_{\ell} =
Im\Delta\delta_{\ell }.
\end{eqnarray}

The total electron elastic-scattering cross section $\sigma_{\rm el}$ is then given by the expression
\begin{eqnarray}
\sigma_{\rm el}=\sum_{\ell =0}^{\infty }\sigma_{\ell},
\end{eqnarray}
where $\sigma_{\ell}$ is the electron elastic-scattering partial cross section:
\begin{eqnarray}
\sigma_{\ell}=\frac{2\pi }{k^2}(2\ell+1)\frac{\cosh{2\mu_{\ell}}-
\cos{2\delta_{\ell}}}{{\rm e}^{2\mu_{\ell}}}.
\label{EqSgmRPAE}
\end{eqnarray}

\section{Results and Discussion: $e + {\rm Ba@C_{60}}$ scattering}

\subsubsection{Preliminary notes}

In the aims of the present paper, the authors focus on electron scattering off Ba@C$_{60}$, as the case study.
This is because the Ba atom is a highly polarizable atom. It is expected to retain its high polarizability under the
C$_{60}$ confinement as well. This provides one with a better opportunity to learn how atomic polarization can alter electron scattering off $A$@C$_{60}$ compared to scattering off the static target.

Furthermore, the study focuses on the electron energy region of up to $\epsilon \approx 3$ eV.
First, at such energies,
the electron wavelength $\lambda > 6$ $\AA$ exceeds greatly
the bond length $D \approx 1.44$ $\AA$ between the carbon atoms in
C$_{60}$. Correspondingly, the incoming electrons will ``see'' the
C$_{60}$ cage as a homogeneous rather than ``granular'' cage. This makes it
appropriate to model the C$_{60}$ cage by a non-granular, i.e., ``smooth'' potential. Moreover, as was shown in Refs.~\cite{Diffuse,Cusp}, a low-energy electron motion in the field of C$_{60}$ is insensitive to details of the
``smooth'' potential, i.e., to whether the potential is the potential with round borders and unparallel walls or simply a square-well potential, as long as  $\lambda \gg D$. This additionally justifies the use of the square-well potential, Eq.~(\ref{SWP}), in the present study.
 Second, correlation/polarization effects, which are of the primary concern of this paper, are expected, as usually, to be most strong primarily at low-energy
 electron collisions with multielectron targets.
 Third, at these low energies, earlier, there were predicted spectacular
confinement-induced resonances in $e + A@{\rm C_{60}}$ scattering \cite{DolmJPB,DolmPRA}, similar to those predicted in $e + {\rm C_{60}}$ scattering \cite{DolmPRA,Winstead,AmBaltC60-}.
The presence of the confining C$_{60}$ cage in this case, as well as in the case of scattering off empty C$_{60}$, results in the emergence of positive quasi-discrete states in the field of $A$@C$_{60}$ or C$_{60}$. When
quasi-discrete states are present, then, in accordance with the known fact, ``... resonance scattering occurs because a positive level with $\ell \neq 0$, though not a true discrete level, is quasi-discrete: owing to the presence of the centrifugal potential barrier, the probability that a particle of low energy will escape from this state to infinity is small, so that the lifetime of the state is long'' \cite{Landau}.

It is interesting to explore how the predicted resonances
in $e + A@{\rm C_{60}}$ scattering can be altered by the effects of polarization of the encapsulated atom $A$ by incident electrons.

Next, the calculations of electron scattering off Ba@${\rm C_{60}}$  and free Ba, performed in the present work in the framework of both SHIFT and SCAT, accounted for
the monopole, dipole, quadrupole, and octupole perturbations of the valence $6s^{2}$ and $5p^{6}$ subshells of free and encaged Ba by incident electrons.
Finally, in view of low values of the electron energies, only the $s$-, $p$-, $d$-, $f$-, and $g$-partial electronic waves have been accounted for in the calculations. The contribution of other partial electronic
waves is negligible, at given energies.

\subsubsection{Results and discussion}

\begin{figure*}[ht]
\center{\includegraphics[width=\columnwidth]{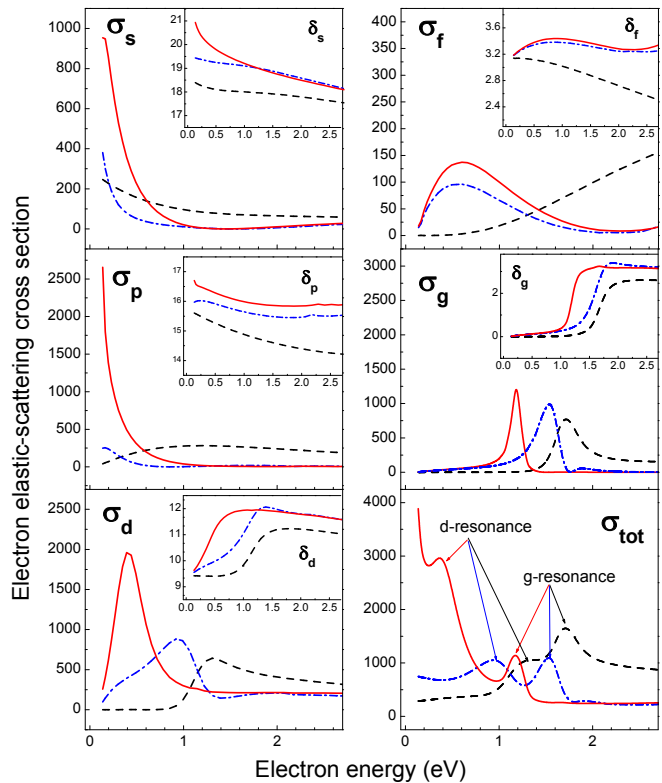}}
\caption{(Color online) Main panels: Calculated partial $\sigma_{\ell}(\epsilon)$ and total $\sigma_{\rm el}(\epsilon)$ cross sections (in units of $a_{0}^2$)
for electron elastic scattering off Ba@C$_{60}$, obtained in the frameworks of the model static HF (dashed line), multielectron SHIFT (dash-dotted line) and SCAT (solid line) approximations.
Insets: Real parts $\delta_{\ell}(\epsilon)$ of the phase shifts $\zeta_{\ell}(\epsilon)$ (in units of radian) calculated in HF (dashed line), multielectron SHIFT (dash-dotted line) and SCAT (solid line) approximations.}
\label{BaC60}
\end{figure*}

Corresponding calculated HF, SHIFT, and SCAT  data for the real parts of phase shifts $\delta_{\ell}(\epsilon)$,  partial $\sigma_{\ell}(\epsilon)$ and total $\sigma_{\rm el}(\epsilon)$ cross sections for
electron elastic scattering off Ba@C$_{60}$ are displayed
in Fig.~\ref{BaC60} (the imaginary parts $\mu_{\ell}$ of the phase shifts, when present, are small at the chosen energies and present little interest for discussion).

As an important finding, one learns from Fig.~\ref{BaC60} that the C$_{60}$ confinement does not at all ``extinguish'' the possibility for the encapsulated Ba atom to be strongly polarized by incident electrons.
On the contrary, the polarization impact is found to affect dramatically both the electron scattering phase shifts and partial $\sigma_{\ell}$ as well as total $\sigma_{\rm el}$  cross sections.
All this is obvious from the comparison of calculated results obtained in the framework of HF, on the one hand, and SHIFT and SCAT, on the other hand, approximations. As another important result, one finds that accounting for only the second-order (with respect to the Coulomb interaction) correlation effects, as in SHIFT, is important but far insufficient for the calculation of low-energy electron scattering off Ba@C$_{60}$. Indeed, the correlation impact
beyond the second-order approximation, i.e., accounted for in the SCAT approximation, is utterly significant -- the lower the energy, the stronger the impact.

For the next, it is both interesting and necessary to bring to the attention of the reader that the \textit{empty} static C$_{60}$  cage can only support the $s$-, $p$-, and $d$-bound states \cite{DolmPRA,Winstead}. In contrast, the ``stuffed'' C$_{60}$, i.e., Ba@C$_{60}$, was found to lose a $s$-bound state, but acquire, instead, a $f$-bound state,
 in the static HF approximation \cite{DolmPRA}. Now, it follows from the present study that by ``unfreezing'' the encapsulated Ba atom, i.e., by making it polarizable, the lost $s$-bound state is returned to the Ba@C$_{60}$ system, and the latter keeps the former $p$-, $d$- and $f$-bound states. Moreover, the Ba@C$_{60}$ system is found to start supporting a second $p$-bound state as well. All this becomes clear by noting the calculated SCAT values of the $s$-, $p$-, $d$-, and $f$-phase shifts at $\epsilon \rightarrow 0$: $\delta_{s}^{\rm SCAT} \rightarrow 7\pi$, $\delta_{p}^{\rm SCAT} \rightarrow 6\pi$, $\delta_{d}^{\rm SCAT} \rightarrow 3\pi$, and $\delta_{f}^{\rm SCAT} \rightarrow \pi$. The zero-energy values of these phase shifts satisfy the generalized version of Levinson theorem for scattering in a potential field \cite{Landau} which (the generalized theorem) could be derived by the direct numerical calculation and written as follows:
\begin{eqnarray}
\left. \delta_{\ell}(\epsilon)\right\vert_{\epsilon \rightarrow 0}\rightarrow (N_{n_{\ell} }+ q_{\ell })\pi.
\label{Levinson}
\end{eqnarray}
Here, $N_{n_{\ell}}$ is the number of closed subshells with given $\ell$ in
the ground-state configuration of an atom, whereas $q_{\ell }$ is the
number of additional \textit{empty} bound states with the same $\ell$ in the field of $A$@C$_{60}$. Given that, for the ground-state of the encapsulated Ba atom,
$N_{n_{s}}=6$, $N_{n_{p}} = 4$, $N_{n_{d}} = 2$, and $N_{n_{f}}= 0$, one finds that, in accordance with the SCAT values of the phase shifts, $q_{n_{s}}=1$, $q_{n_{p}} = 2$, $q_{n_{d}} = 1$, and $q_{n_{f}}= 1$.
This translates into one $s$-, two $p$-, one $d$-, and one $f$-bound states supported (one at a time) by Ba@C$_{60}$. Note that calculated SHIFT phase shifts, in contrast to SCAT data, point to the existence of
neither $s$- nor second $p$-bound state in the field of Ba@C$_{60}$. This discrepancy between the calculated SHIFT and SCAT data speaks, in general, to the importance of a fuller account of electron correlation in the
calculation of $e + $A$@{\rm C_{60}}$ scattering.

The discovered in the framework of SCAT emergence of a $s$-bound state and a second $p$-bound state in the field of Ba@C$_{60}$ has profound consequences for both the corresponding partial and total electron-scattering cross sections. Namely, because the phase shift $\delta_{s}^{\rm SCAT}$,  on the way to its value of $7\pi$ at $\epsilon=0$,  passes through the value of modulo $\pi/2$ (at $\epsilon \approx 0.23$ eV),
$\sigma_{\rm s}^{\rm SCAT}$ becomes large, at lower energies, in contrast to the predictions obtained with the help of inferior HF and SHIFT. Similarly, the increase of $\delta_{\rm p}^{\rm SCAT}$ to $6\pi$ at $\epsilon =0$ results in large
$\sigma_{\rm p}^{\rm SCAT}$ as well, at low energies, again, in contrast to the predictions by HF and SHIFT.

Talking about the $d$- and $g$-partial cross sections, one can see that each of them is dominated by the strong resonance.  Its emergence  clearly follows from the behavior of the $d$- and $g$-phase shifts. Indeed, the phase shifts  first tend to increases towards modulo $\pi$ with decreasing energy, but, because the field turns out to be not strong enough, before that value is reached, they
sharply decrease, passing through the value of modulo $\pi/2$. This is a typical behavior of phase shifts upon electron scattering on quasibound states \cite{Newton}.
Next, note how the resonances in the $d$- and $g$-partial cross sections become higher, narrower, and shift towards lower energy as more correlation is accounted for in the calculation (compare
calculated HF versus SHIFT versus SCAT results). We thus find that the inclusion of more correlation into the calculation of $e + {\rm Ba@C_{60}}$ scattering increases the strength of the
field of the Ba@C$_{60}$-scatterer, since the above noted changes in the resonances are typical for electron scattering in a field of increasing strength \cite{Newton}.

Next, note that no $f$-resonance emerges in $e + {\rm Ba@C_{60}}$ scattering calculated in either of the three approximations utilized in this paper. This is in contrast to electron elastic-scattering off empty static
C$_{60}$. There, the sharp $f$-resonance was predicted to emerge at low energies \cite{DolmPRA,Winstead,AmBaltC60-} (this is the extremely narrow resonance near zero plotted in Fig.~\ref{C60}).  The reason for the absence of the $f$-resonance in  $e + {\rm Ba@C_{60}}$ scattering is interesting. It is directly associated with that a noticeable part of the valence electronic charge-density of encapsulated Ba is drawn into the C$_{60}$ shell \cite{DolmPRA}.
Therefore, the field inside the ``stuffed'' C$_{60}$ becomes more attractive than in empty C$_{60}$. It turns the $f$-state into a bound state, thereby eliminating the emergence of a $f$-resonance
in $e + {\rm Ba@C_{60}}$ scattering. Now, as it has been shown in the paragraph above,  the inclusion of correlation into the calculation of $e + {\rm Ba@C_{60}}$ scattering increases the field of Ba@C$_{60}$. Therefore,  the $f$-state remains bound. This is why the $f$-resonance does not take place in $e + {\rm Ba@C_{60}}$ scattering even if polarization of the encapsulated Ba atom by incident electrons is accounted for in the calculation. Our general prediction is that there will be no $f$-resonances on quasibound states in electron scattering off any $A$@C$_{60}$ system in case where there is a noticeable transfer of electronic charge-density from the encapsulated atom to the C$_{60}$ shell.

\begin{figure}[ht]
\center{\includegraphics[width=\columnwidth]{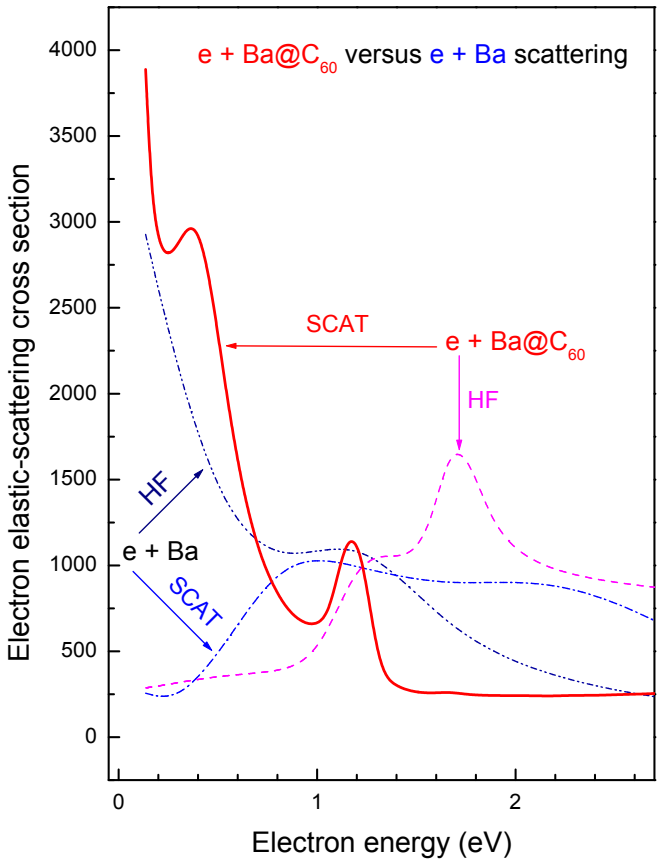}}
\caption{(Color online) Calculated total electron elastic-scattering cross sections $\sigma_{\rm el}(\epsilon)$ (in units of $a_{0}^2$)
for electron scattering off Ba@C$_{60}$, obtained in the frameworks of the model static HF (dashed line) and multielectron SCAT (solid line) approximations,
as well as off free Ba (HF, dash-dot-dot; SCAT, dash-dot), as marked.}
\label{BaVSBaC60}
\end{figure}

Furthermore, it is interesting to compare how much differently polarization of the Ba atom by incident electrons affects electron elastic scattering off free Ba versus
Ba@C$_{60}$. The corresponding calculated HF and SCAT total electron elastic-scattering cross sections are depicted in Fig.~\ref{BaVSBaC60}.

The calculated data reveal a spectacular difference between the role of polarization in electron scattering off Ba and Ba@C$_{60}$. Namely,
it appears that the effects of polarization in $e + {\rm Ba@C_{60}}$ scattering act oppositely to the effects in $e + {\rm Ba}$ scattering. Thus, whereas
$\sigma_{\rm el}^{\rm SCAT}(e +{\rm Ba}) \ll \sigma_{\rm el}^{\rm HF}(e +{\rm Ba})$ at $\epsilon \alt 1$ eV, the situation
is exact opposite for $e + {\rm Ba@C_{60}}$ scattering in about the same energy region:
 $\sigma_{\rm el}^{\rm SCAT}(e +{\rm Ba@C_{60}}) \gg \sigma_{\rm el}^{\rm HF}(e +{\rm Ba@C_{60}})$.
 Alternatively, whereas $\sigma_{\rm el}^{\rm SCAT}(e + {\rm Ba}) \gg \sigma_{\rm el}^{\rm HF}(e + {\rm Ba})$
at $\epsilon \agt 1.4$ eV, one observes that $\sigma_{\rm el}^{\rm SCAT}(e + {\rm Ba@C_{60}}) \ll \sigma_{\rm el}^{\rm HF}(e + {\rm Ba@C_{60}})$ in there.
It is, thus, found in the present study that the effects of atomic polarization in electron scattering off the free and encapsulated inside C$_{60}$ atoms
may follow opposite routes. This is an interesting observation.

Lastly, note that there are energy regions, specifically,
$0.8 \alt \epsilon \alt 1.1$ eV and $\epsilon \agt 1.2$ eV,  where
$\sigma_{\rm el}^{\rm SCAT}(e + {\rm Ba@C_{60}}) \ll \sigma_{\rm el}^{\rm SCAT}(e + {\rm Ba})$. This means that the gas-medium of big-sized $A@{\rm C_{60}}$s
can be more transparent to incident electrons than the gas-medium of smaller-sized isolated atoms $A$ themselves.
This counter-intuitive effect was earlier unveiled in Ref.~\cite{DolmJPB} in the framework of the static HF approximation, but appears to
retain its place even if the encapsulated atom is polarizable, as is shown in the present paper.

\section{Conclusion}

The present work has provided a deeper insight  into
possible features of low-energy electron elastic scattering  off
$A$@C$_{60}$ fullerenes.  This has been
achieved by studying the dependence of $e + {\rm Ba@C_{60}}$ elastic scattering with account for polarization
 of encapsulated Ba by incident electrons. It has been demonstrated that the polarization effect
results in dramatic differences between electron scattering off Ba@C$_{60}$ evaluated with and without inclusion of polarization into the calculation.
It has been found that a fuller account for correlation effects in $e + A@C_{60}$ scattering is utterly important. Furthermore, it has been unraveled in the present study
that the impact of polarization on electron scattering off $A$@C$_{60}$ may be both qualitatively and quantitatively different than that
in the case of
electron scattering by the free atom $A$. For instance, it has been demonstrated that  where polarization significantly enhances the $e + {\rm Ba@C_{60}}$
scattering cross section, it significantly diminishes the $e + {\rm Ba}$ scattering cross section and vice verse. This leads to the possibility for electron scattering
off $A$@C$_{60}$ to become significantly weaker than in the case of electron scattering by the isolated atom $A$, in certain energy regions. This counter-intuitive effect
has been found to be stronger and occur in a broader energy region than when polarization is ignored.

Lastly, the present study provides researchers with
background information which is useful for future studies of electron scattering by $A$@C$_{60}$, particularly aimed
at elucidating of a possible significance of a simultaneous polarization of both the C$_{60}$ cage and encaged atom by incident electrons. This
will make the $A$@C$_{60}$ more attractive, so that predicted in the present study features of $e + A@{\rm C_{60}}$ may appear at different energies, or disappear
at all, some actual bound states may be converted to resonances, etc. Such effects, however, are subject to an
independent study.

\section{Acknowledgements}

V.K.D. acknowledges the support by NSF Grant No.\ PHY-1305085.

\section*{References}

\end{document}